\documentstyle[psfig]{mn}

\def\simless{\mathbin{\lower 3pt\hbox
   {$\rlap{\raise 5pt\hbox{$\char'074$}}\mathchar"7218$}}}   
\def\simgreat{\mathbin{\lower 3pt\hbox
   {$\rlap{\raise 5pt\hbox{$\char'076$}}\mathchar"7218$}}}   
\def\etal{{\rm et al.}}

\def\solm{{M_\odot}}

\def\tff {t_{\rm ff}}

\def\ms {M_*}
\def\msi {M_*(0)}
\def\mgas {M_{\rm gas}}
\def\mstars {M_{\rm stars}}

\def\Rroche {R_{\rm tidal}}
\def\Racc {R_{\rm acc}}
\def\vel {V_{\rm rel}}
\def\macc {\dot M_*}
\def\menc {M_{\rm enc}}
\def\Rclus {R_*}
\def\Rclus {R}
\def\Rbh {R_{\rm BH}}

\ifnfsstwo

\fi

\ifnfssone
  \newmathalphabet{\mathit}
    \addtoversion{normal}{\mathit}{cmr}{m}{it}
    \addtoversion{bold}{\mathit}{cmr}{bx}{it}

\fi

\ifoldfss

\fi

  \makeatletter
  \ifx\CUP@mtlplain@loaded\undefined  
  \makeatother  
    
  \else
    
  \fi

\loadboldmathitalic 
\loadboldgreek      
  
\makeatletter
\ifx\CUP@mtlplain@loaded\undefined  
  \newfont\bit{cmbxti10 at 9pt}
\else
  \newfont\bit{mtbxti10 at 9pt}
\fi
\makeatother  
 
\title[Accretion and the IMF] {Accretion in stellar clusters and the IMF}
  \author[I. A. Bonnell \etal]  {I. A. Bonnell$^1$, C. J. Clarke$^2$, M. R. 
  Bate$^2$ and J. E. Pringle$^2$
 \\ $^1$ School
of Physics and Astronomy, University of St Andrews, North Haugh, St
Andrews, Fife, KY16 9SS. \\
$^2$ Institute of Astronomy, Madingley
Road, Cambridge CB3 0HA.} \date{\today}
\pagerange{\pageref{firstpage}--\pageref{lastpage}}
\pubyear{    }

\def\LaTeX{L\kern-.36em\raise.3ex\hbox{a}\kern-.15em
    T\kern-.1667em\lower.7ex\hbox{E}\kern-.125emX}

\begin{document}

\label{firstpage}

\maketitle

\begin{abstract}
We present a simple physical mechanism that can account for the
observed stellar mass spectrum for masses $\ms \simgreat 0.5 \solm$.
The model depends solely on the competitive accretion that occurs in
stellar clusters where each star's accretion rate depends on the local
gas density and the square of the accretion radius. In a stellar
cluster, there are two different regimes depending on whether the gas
or the stars dominate the gravitational potential.  When the cluster
is dominated by cold gas, the accretion radius is given by a
tidal-lobe radius. This occurs as the cluster collapses towards a
$\rho\propto R^{-2}$ distribution. Accretion in this regime
results in a mass spectrum with an asymptotic limit of $\gamma=-3/2$
(where Salpeter is $\gamma=-2.35$).  Once the stars dominate the
potential and are virialised, which occurs first in the cluster core,
the accretion radius is the Bondi-Hoyle radius. The resultant mass
spectrum has an asymptotic limit of $\gamma=-2$ with
slightly steeper slopes ($\gamma\approx-2.5$) if the stars are already mass-segregated. 
Simulations of
accretion onto clusters containing 1000 stars show that as expected,
the low-mass stars accumulate the majority of their masses during the
gas dominated phase whereas the high-mass stars accumulate the
majority of their massed during the stellar dominated phase. This
results in a mass spectrum with a relatively shallow $\gamma\approx
3/2$ power-law for low-mass stars and a steeper, power-law for
high-mass stars $ -2.5\simless\gamma\le -2$.   This competitive accretion model also
results in a mass segregated cluster.

\end{abstract}


\section{Introduction}

One of the most fundamental unsolved problems in astronomy is the
origin of the distribution of stellar masses, the so-called Initial
Mass Function (IMF).  Significant observational work has gone into
establishing the exact form of the mass spectrum (c.f. Scalo~1986;
Kroupa, Tout \& Gilmore~1990; Kennicut~1998) and investigating it's
universality (Kennicut~1998; Leitherer~1998;
Scalo~1998). Unfortunately this work has not led to an increase in our
understanding of how the mass spectrum originates.  This shortcoming
is intrinsically linked to our lack of understanding of the star
formation process.  Thus, theories of the IMF have generally involved
sufficient free-parameters to ensure that they can reproduce the
general form of the mass spectrum but also, unfortunately, ensuring
that they do not help us discern the physics involved.

Observations of star forming regions have clearly shown that most
stars are formed in clusters (e.g. Lada \etal~1991; Lada, Strom and
Myers 1993; Meyer \& Lada~1999; Clarke, Bonnell \&
Hillenbrand~2000). Furthermore, the individual stellar masses found in
these clusters span the observed range of stellar masses and are
consistent with being drawn from the field star IMF
(Hillenbrand~1997; Hillenbrand \& Carpenter~2000). Thus, the basic physics which results in the
observed mass spectrum is present in this fundamental building block
of the star formation process: the young stellar cluster. Competitive
accretion in gas-rich stellar clusters results in a spectrum of
stellar masses (Bonnell \etal~1997,~2000) and in this {\sl paper} we investigate
whether this process can lead to the observed IMF.

There have been several theories advanced to explain the observed
distribution of stellar masses (cf. Clarke~1998; Meyer
\etal~2000). These theories generally involve either fragmentation or
accretion plus feedback as the basic physics in setting the stellar
masses.  Fragmentation of a giant molecular cloud can set the stellar
mass distribution if the GMC clump-mass distribution is, for some
reason, of the same form as the IMF (Elmegreen~1997), or if feedback
changes the basic fragment (Jeans) mass (Silk~1977).  Alternatively,
accretion onto a small isolated protostellar core can set the final
masses but this requires a halting mechanism (eg feedback) that really
determines the mass distribution (Adams \& Fattuzzo~1996). The main
criticisms of these theories are that a) they rely on additional
processes (e.g. feedback or an initial mass distribution) and b) that
they neglect the fact that the stars are not formed in isolation but
in clusters and therefore can interact during their
formation. Furthermore, they are generally unable to explain why
clusters are initially mass-segregated (eg, Hillenbrand \&
Hartmann~1998; Bonnell \& Davies~1998) which should be a 
requirement of any theory that attempts to explain the IMF. In this
{\sl paper}, we discuss how competitive accretion in clusters can set
the distribution of stellar masses and result in a mass-segregated cluster.

\section{Accretion in Clusters}

Young stellar clusters are generally found to contain significant
amounts of gas. In fact, the total gas mass is typically many
times the mass in stars (Lada~1991). This gas is a remnant of the star
formation process and can be understood in terms of the inefficiency
of fragmentation where the majority of a system's mass is not
immediately included in the fragments but is accreted by them over
longer time periods (Larson~1978; Boss~1986; Bonnell \& Bastien~1992;
Burkert \& Bodenheimer~1993; Klessen, Burkert \& Bate~1998).  This
occurs due to the non-homologous nature of gravitational collapse
(Larson~1969) so that the amount of mass that has reached the region
where and when fragmentation occurs, is small. Thus, the initial
fragment mass should generally be small, $M<<\solm$. Furthermore, in a
cluster environment, the fragment mass, if it is proportional to the
Jeans mass, should be smallest in the centre of the cluster where the
density is highest, in direct contradiction to the observed mass
segregation (Bonnell \& Davies~1998; Bonnell \etal~1998).  It is thus
unlikely that the mass distribution is solely determined by the fragmentation
process.

Accretion of the residual gas in clusters can play a crucial role in
setting the final stellar masses (Zinnecker~1982; Larson~1992). As all
the stars are embedded in the same envelope of gas, they compete for
the gas as they move through the cluster. This competition results in
unequal mass accretion rates and hence a mass spectrum

Simulations of accretion in small stellar clusters (Bonnell \etal~1997)
have shown that even if the stars have equal initial masses, they
accrete unevenly with those stars near the centre accreting more than those
near the outer parts of the cluster. This dependency of the accretion
rates on position in the cluster is even clearer in larger clusters 
($N=100$; Bonnell \etal~2000) where accretion rates can vary by
over an order of magnitude from the outside to the cluster centre
when the cluster is centrally-condensed.
The physics behind this differential accretion is the effect of the
cluster potential in funneling material down to the centre plus the individual
star's ability to capture this material. Initially uniform clusters
also accrete differentially as they evolve towards a centrally-condensed configuration, due to the local competition
between stars. The final cluster displays a significant degree
of mass segregation due to the differential accretion (Bonnell \etal~2000).

A star's accretion rate depends on it's cross section for accreting material, $\pi \Racc^2$, the gas density through which it moves, $\rho$ and the velocity
with which it moves relative to the gas, $\vel$:
\begin{equation}
\macc = \pi \rho \vel \Racc^2.
\end{equation}
The question is what is the accretion radius, $\Racc$. An isolated
star of mass $\ms$, moving in a uniform medium with velocity $\vel$
relative to the gas with sound speed $c_s$, has an accretion radius given by
Bondi-Hoyle accretion where 
\begin{equation}
\Rbh = 2G\ms/(\vel^2 + c_s^2)
\end{equation}
(e.g. Bondi \& Hoyle~1944; Ruffert~1996). On the other hand, in a
cluster environment where cold gas dominates the potential, 
 the accretion radius can be described by the star's
tidal-lobe radius.
\begin{equation}
\Rroche \approx 0.5  \left({\ms \over \menc}\right)^{1\over 3} \Rclus,
\end{equation}
where $\ms$ is the star's mass, $\menc$ is the cluster mass interior
to the star's position in the cluster $\Rclus$ (e.g. Paczynski~1971;
Frank, King \& Raine 1985). Material inside this tidal-lobe is more attracted to the
star whereas outside this lobe the acceleration is dominated by the
cluster potential.

Simulations of accretion in gas-dominated stellar clusters have
shown that the accretion is well modelled by a tidal-lobe 
radius as long as the gas dominates the gravitational potential
(Bonnell \etal~2000).
When this is the case and the gas is unsupported, as expected in order
for the cluster to have fragmented, both the gas and the stars
collapse towards the cluster centre. The stars are impeded from
virialising due to the changing potential, and the gas accretion. 
Therefore  the gas and stellar velocities are locally 
correlated such that the relative gas velocity is small. This
results in a large $\Rbh$ such that $\Rbh > 
\Rroche$ and therefore the smaller $\Rroche$ is the appropriate accretion radius.

In contrast, once the stars dominate the 
gravitational potential, which occurs first in the core of the cluster
due to the higher accretion rates there and
the sinking of the more massive stars, 
the stars virialise and their velocities
relative to the gas become large. When this occurs the Bondi Hoyle
radius becomes much smaller due to the large gas velocities such that
$ \Rbh < \Rroche$. At this point, the Bondi-Hoyle radius is a better description
of the accretion radius (Bonnell \etal~2000).

\section{The Initial Mass Function}

In order to use the above formulations of the accretion rate to
determine the resultant mass spectrum, we need a basic model of a stellar
cluster.  The cluster is initially gas dominated in order to
provide the mass reservoir for accretion. Observations of the youngest
clusters estimate that up to 90 per cent of the mass is in gas
(Lada~1991).  Such systems are unlikely to be in equilibrium, and
regardless of their initial configurations, will collapse towards a 
\begin{equation}
\label{isogas}
\rho \propto R^{-2}
\end{equation}
 density distribution (Larson~1969; Hunter~1977). The
stars will also collapse, even if initially virialised, due to the
change in the potential. Thus we assume that the stars follow the same distribution:
\begin{equation}
\label{nofR}
n \propto R^{-2}
\end{equation}
 This profile  is a good approximation to the stellar distribution of most clusters
(Binney \& Tremaine~1987) including the youngest clusters that have
been studied such as the Orion Nebula cluster (Hillenbrand \& Hartmann~1998). Furthermore, it is the
expected distribution produced by a violent relaxation of a 
non-equilibrium purely stellar cluster (Lynden-Bell~1967).

The gas distribution will be modified by the accretion and hence
removal of material. The accretion rates are
largest in the centre and hence it is there that they will first
affect the density profile. This gas removal destabilises the equilibrium
of the isothermal sphere allowing gas to infall from further out in the
cluster. In the case of infall onto a point-mass, the density profile then takes on the form (Hunter~1977; Shu~1977;
Foster \& Chevalier~1993)
\begin{equation}
\label{gasacc}
\rho \propto R^{-{3\over 2}}.
\end{equation}
We thus assume an initial density profile of an isothermal sphere
(equation~\ref{isogas}) which evolves towards a profile composed
of the shallower, equation(\ref{gasacc}) profile in the central regions
while the outside maintains the isothermal sphere profile. The 
transition
between these two regimes occurs when gas accretion has increased the
stellar masses sufficiently that the stars dominate the gravitational
potentialin the cluster centre. We
discuss the asymptotic limits to the accretion and mass spectrum
in these two cases below.

\subsection{Gas dominated potentials}

In addition to the gas density we require estimates of the
relative gas velocity and of the accretion radius.
Where gas dominates the potential 
($\rho \propto R^{-2}$), 
the accretion radius is given by the tidal-lobe radius (Bonnell \etal~2000)
\begin{equation}
\Racc \approx \Rroche.
\end{equation}
In order to evaluate the tidal-lobe radius we need the mass 
enclosed at a particular radius in the cluster, $\Rclus$, 
\begin{equation}
\menc = \int_0^{\Rclus} 4 \pi \rho r^2 dr,
\end{equation}
which in
a $\rho\propto R^{-2}$ distribution is
\begin{equation}
\menc \propto \Rclus.
\end{equation}

Simulations of accretion in stellar clusters have shown that in 
gas-dominated clusters, the 
relative gas velocity is generally subsonic as both the stellar and gas 
velocities are dominated by the acceleration of the cluster 
potential (Bonnell \etal~2000). 
Alternatively, if the stars are virialised then the
relative gas velocity is the velocity dispersion, 
\begin{equation}
\vel \approx \sqrt{G \menc \over R},
\end{equation}
Thus, if the gas distribution follows equation(\ref{isogas}), then we have
\begin{equation}
\vel = constant.
\end{equation}
In either case the relative gas velocity
is independent of position in the cluster.

We can now use the above to consider how accretion in a stellar
cluster results in a mass spectrum of the form
\begin{equation}
dN \propto \ms^{\gamma} d\ms
\end{equation}
where $\gamma=-2.35$ is the Salpeter IMF.
First of all, we note that from equation(\ref{nofR}) the 
number of stars at a given radius is constant:
\begin{equation}
\label{stelldist}
dN = n(R) 4 \pi R^2 dR \propto {\rm const} \times dR.
\end{equation}
Starting from the general equation for the accretion rates,
\begin{equation}
\label{acceqn}
\macc  = \pi \rho \vel \Racc^2,
\end{equation}
and adapting it for the case where gas dominates the potential
such that $\rho\propto R^{-2}$ and $\Racc \approx \Rroche$, 
then we have
\begin{equation}
\label{tidmacc}
\begin{array}{ll}
\macc & \propto g(t) \Rclus^{-2} \times \left(\left[\ms\over\menc\right]^{1\over 3}
\Rclus\right)^2 \\
 & \propto g(t) \left( \ms\over\menc\right)^{2\over 3} \\
& \propto g(t) \left( \ms\over\Rclus\right)^{2\over 3}.
\end{array}
\end{equation}
The term $g(t)$ includes all time-dependence of the homologously
evolving cluster and $\Rclus$ is the initial stellar position in the
cluster.

Integrating equation(\ref{tidmacc}) with respect to time, $t$, and in the
limit where accretion dominates the final mass, we get
\begin{equation}
\label{tidmsvsrad}
\ms \propto  \Rclus^{-2} h(t)^3,
\end{equation}
where $h(t) = \int g(t) dt$.
Thus, at a time $t_{\rm final}$, when the accretion is halted, we have
\begin{equation}
\label{massvsrad1}
R \propto \ms^{-{1\over 2}}
\end{equation}
and thus
\begin{equation}
\label{deltaR}
dR \propto \ms^{-{3\over 2}} d\ms.
\end{equation}
This results in a mass spectrum (equation~\ref{stelldist})
\begin{equation}
dN \propto \ms^{-{3\over 2}} d\ms.
\end{equation}

In general, the $\rho\propto R^{-2}$ profile will not extend right to
the centre of the cluster. Instead, the central regions must have a
flatter profile to stop the density from becoming infinite.  A collapsing
isothermal gas cloud has a $\rho\propto R^{-2}$ density profile in the outer
regions and an ever decreasing central region with near uniform
density (Larson~1969). Stars in this region will see a shallower
density profile which will affect their accretion rates
(equation~\ref{acceqn}). Adapting equation(\ref{tidmacc}) for a density
profile $\rho\propto R^{-\alpha}$ while maintaining a uniform $\vel$
gives
\begin{equation}
\label{massvsrad}
\ms \propto R^{- \alpha}
\end{equation}
and analogous to equation(\ref{deltaR}),
\begin{equation}
dR \propto \ms^{-{1\over \alpha} - 1} d\ms. 
\end{equation}
Taking the stellar
density to be proportional to the gas density, $n(R) \propto \rho (R)$, 
equation(\ref{stelldist}) becomes
\begin{equation}
\label{gausimf}
\begin{array}{ll}
dN & = n(R) 4 \pi R^2 dR \propto R^{2-\alpha} dR, \\
& \propto \ms^{-{2\over\alpha}} \ms \ms^{-{1\over \alpha} - 1} d\ms, \\
& \propto \ms^{-{3\over\alpha}} d\ms,
\end{array}
\end{equation}
and $\gamma$ is now $\gamma= -3/\alpha$.
\begin{figure}
\psfig{{figure=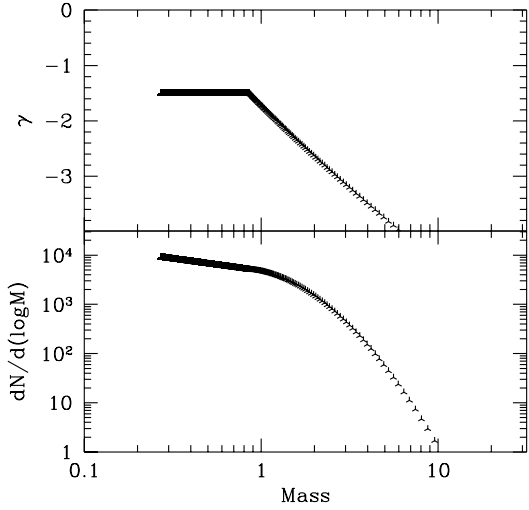,width=3.3truein,height=3.3truein,rwidth=3.0truein,rheight=3.0truein}}
\caption{\label{gausimffig} The resultant IMF as derived from
equation(\ref{gausimf}), is plotted in log mass bins for a gas
dominated cluster with $\rho \propto R^{-2}$ on the outside and a
Gaussian central density profile (lower panel). The masses are scaled
such that the average mass is $0.4 \solm$. The upper panel shows the
corresponding value of $\gamma$ where $dN = \ms^{\gamma} d\ms$. 
The lower-mass stars have
a $\gamma=-3/2$ ($\Gamma=-1/2$) mass spectrum (where Salpeter is
$\gamma=-2.35$) due to the $\rho \propto R^{-2}$ whereas the more
massive stars have an ever steepening mass spectrum due to the central
Gaussian density profile (upper panel).}
\end{figure}
From equation(\ref{gausimf}) we see that when the density profile
is shallower than $R^{-2}$, the resulting mass spectrum is steeper than 
$\ms^{-3/2}$. Figure~\ref{gausimffig} plots the resulting IMF using
equation(\ref{gausimf}) and a density profile that combines an
outer region with a $\rho\propto R^{-2}$ profile with an inner
Gaussian density profile (fitted onto the outer region where it
has the same slope). The mass spectrum is plotted as a function of
the logarithm of the mass such that the slope is $\Gamma = \gamma + 1$. 
The low-mass stars that form in the outer region
have the $\gamma = {-3/2}$ mass spectrum whereas the higher-mass
stars formed in the inner regions have an ever steeper mass spectrum
due to the central Gaussian density profile that converges to uniform
density. 
It is also worth noting that if the collapse gives a density profile
that is steeper than $R^{-2}$ as was found in the outer regions of
the collapse of a Bonnor-Ebert spheroid (Foster \& Chevalier~1993), then
this translates (equation[\ref{gausimf}]) into a shallower IMF
for the low-mass stars that are in this region.

\subsection{Stellar dominated potentials}

In contrast to a gas dominated system,
when the cluster potential is dominated by the stars
and they are virialised, 
the appropriate accretion 
radius is the Bondi-Hoyle radius (Bonnell \etal~2000).
The Bondi-Hoyle radius models better the accretion radius as 
the gas and stellar velocities are no longer correlated.
The gas is primarily
infalling from further out in the cluster onto the stellar
dominated core. When infall is onto a point mass, the
resulting velocity profile is
\begin{equation}
\vel \propto R^{-1/2}
\end{equation}
 which assuming a constant mass infall rate with radius implies a 
density profile of the form  of equation(\ref{gasacc}),
\begin{equation}
\rho\propto R^{-3/2}.
\end{equation}
In the case of accretion onto a stellar cluster, the distributed
mass and the presence of multiple accretion sources will modify
these profiles. We can generalise them as
\begin{equation}
\vel \propto R^{-\eta},
\end{equation}
and
\begin{equation}
\rho\propto R^{-\zeta},
\end{equation}
under the constraint that the mass infall, $4 \pi R^2 \rho \vel$ decreases with
decreasing radius, $\eta + \zeta \le 2$.
we have
\begin{equation}
\label{maccbh}
\begin{array}{ll}
\macc & \propto \vel \rho  \Rbh^2 \\
& \propto \vel \rho  \left(\ms\over \vel^{2}\right)^2 \\
& \propto \rho  {\ms^{2}\over \vel^{3}} \\
& \propto g(t) \Rclus^{-\zeta}  \ms^{2} \Rclus^{3 \eta}\\
& \propto g(t) \ms^{2} \Rclus^{3\eta - \zeta}.
\end{array}
\end{equation}
As above, $g(t)$ includes the time-dependence of the homologously
evolving cluster and $\Rclus$ is the initial stellar position
in the cluster.
In the case where $3\eta - \zeta = 0$ as occurs for accretion onto
a point mass, the accretion rates are  
independent of position in the cluster.

Assuming that the cluster maintains the same form ($3\eta - \zeta\approx$ constant), 
solutions to equation(\ref{maccbh}) are
\begin{equation}
\label{maccbhsol}
\ms = {\msi \over 1 - \beta \msi h(t)}
\end{equation}
where $\msi $ is the initial stellar mass, $h(t) = \int g(t) dt$, 
and $\beta$ 
is basically the accretion clock which 
depends on the gas content and on $\Rclus^{3\eta-\zeta}$ but not on time.
The primary difference between this case and the preceding tidal-lobe
accretion is the higher dependency on the stellar mass such that
the initial mass is always important. This is the case for
any accretion rate which depends on the stellar mass to a power
greater than one.

The resulting mass spectrum from such an accretion rate was studied by
Zinnecker~(1982) and we follow his derivation here.  As there is a
direct correlation between initial and final mass, then there is a
direct mapping from the initial mass spectrum to the final one,
\begin{equation}
\label{imfmap}
F(\ms) d\ms = F(\msi) d\msi.
\end{equation}
Inverting equation(\ref{maccbhsol}) and inserting it into
equation(\ref{imfmap}), as long as $\beta$ is independent of the 
initial masses, we have
\begin{equation}
\begin{array}{ll}
F(\ms) & = F(\msi) {\left({d\ms \over d\msi}\right)}^{-1} \\
&= F\left({\msi}\right) \left({d \over d\msi} 
{\msi \over 1 - \beta \msi h(t)}\right)^{-1}, \\
 &=  F\left(\msi\right) \left({1 - \beta 
 \msi h(t)}\right)^2, \\
 &= F\left({\ms \over 1 + \beta \ms h(t)}\right) \left({\msi \over 
 \ms}\right)^{2}.
\end{array}
\end{equation}
In the limit $\beta \ms h(t) >> 1$ ($\ms >> \msi$) this gives
\begin{equation}
F(\ms) \propto \ms^{-2},
\end{equation}
and thus 
as
long as there is an initial (small) range of stellar masses, the
asymptotic limit of the 
mass spectrum is 
\begin{equation}
dN \propto \ms^{-2} d\ms.
\end{equation}

\begin{figure}
\psfig{{figure=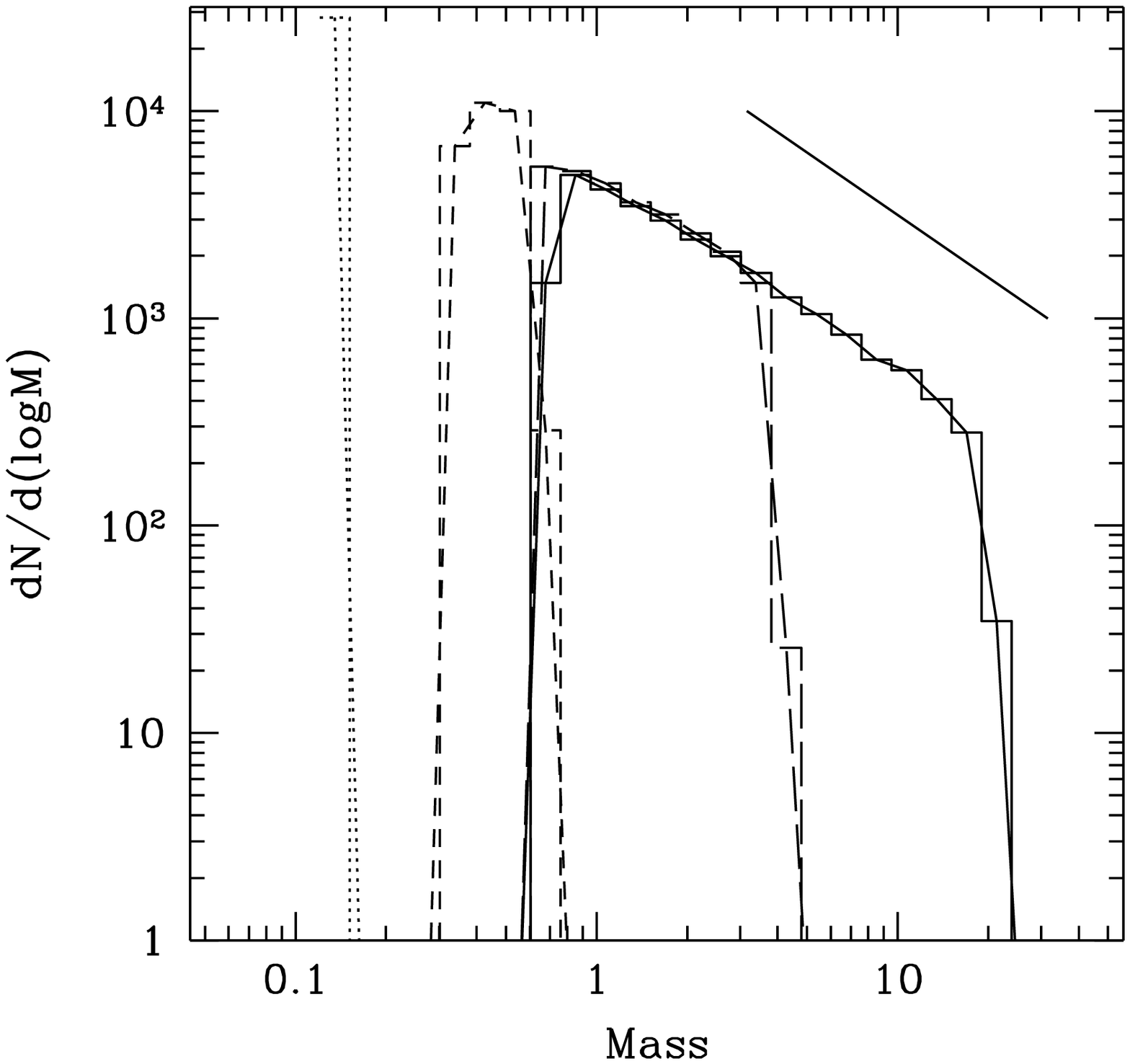,width=3.3truein,height=3.3truein,rwidth=3.0truein,rheight=3.0truein}}
\psfig{{figure=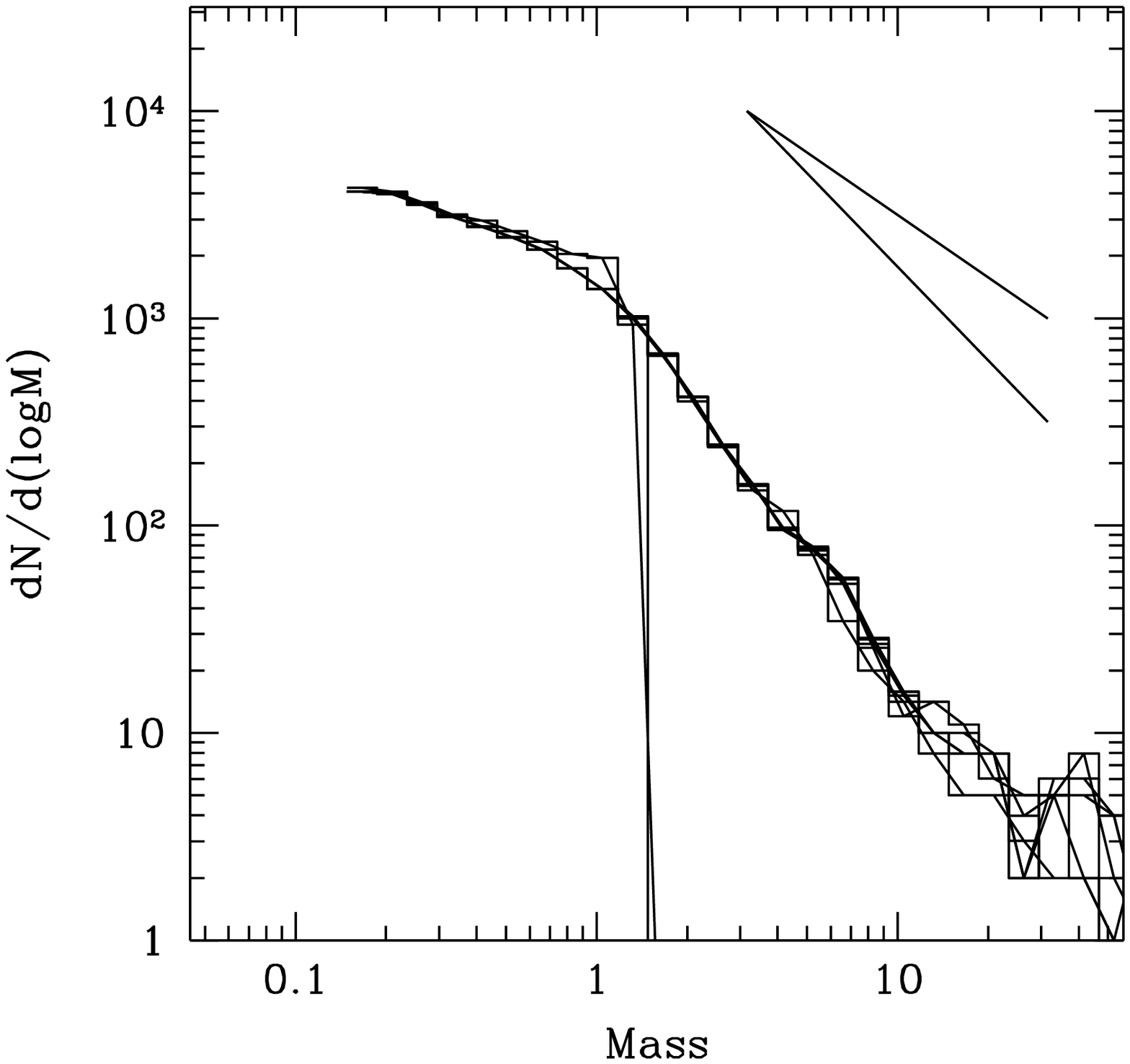,width=3.3truein,height=3.3truein,rwidth=3.0truein,rheight=3.0truein}}
\caption{\label{bhimf3o2} Simulated  IMFs for a combined 28 clusters containing
1000 stars each are
plotted in log mass bins. The clusters are assumed to have stellar
dominated potentials and accrete according to  either 
equation(\ref{maccbh}) with no spatial dependence (upper panel) or
equation(\ref{maccbhseg}) for an initially mass-segregated cluster
(lower panel). The stars are in a $n \propto R^{-2}$ configuration
and are initially a delta function in mass (upper panel) or the expected
$\gamma=-3/2$ ($\Gamma=-1/2$) mass
spectrum from tidal-lobe accretion (where Salpeter is $\gamma=-2.35$).
The higher-mass stars develop the steeper $\gamma = -2$ and $\gamma=-5/2$ ($\Gamma=-3/2$) mass
spectra due to the accretion. The solid lines demonstrate the $\gamma=-2$ and $\gamma=-2.5$ slopes.}
\end{figure}

This is somewhat complicated if there is a correlation between the 
initial masses and radii in the cluster. In this case $\beta$ is 
dependent
on $\msi$ and we cannot derive an analytical expression for the 
resulting mass 
spectrum. In the case where the incoming mass distribution is from
the tidal-lobe accretion, then from \S 3.1 we have
an expression for the initial mass-radius correlation 
(equation~\ref{massvsrad1}). Using this, we can generate
a mass spectrum from accretion for a cluster in a $n\propto R^{{-2}}$ configuration.
Numerical simulations of accretion in stellar dominated clusters
have values
of $\eta = 1/3$ and $\zeta = 3/2$ (Bonnell \etal~2000).
Equation~(\ref{maccbh}) then becomes
\begin{equation}
\label{maccbhseg}
\macc \propto \ms^{2} \Rclus^{-{1\over 2}}.
\end{equation}
Figure~\ref{bhimf3o2} shows the resultant IMF for clusters accreting
from an initial delta function mass distribution and those with an
initially $\ms^{-3/2}$ mass spectrum that accretes according to
equation~(\ref{maccbhseg}). In the former case, the stars, in a
$n\propto r^{-2}$ distribution, accrete with a spatially-independent
accretion rate ($\macc \propto \ms^2$). These clusters develop an
$\ms^{-2}$ mass spectrum. In contrast, with the initial mass
segregation, the IMF develops into a steeper ($\gamma \approx -2.5$)
IMF as the stars accrete in the stellar dominated regime. This is
somewhat steeper than the analytical $\ms^{{-2}}$ mass spectrum due to
the initial correlation between stellar mass and position in the
cluster.

We expect stellar dominated clusters to accrete towards
a mass spectrum with asymptotic limits of $-2.5 \simless \gamma \le -2$.
This mass spectrum is expected for the higher mass stars as it is they
that are located in the centre of the cluster and have attained a
higher mass due to their previous accretion in the gas dominated
regime. Clusters that have mass distributions that are uncorrelated with 
cluster position going into the stellar
dominated phase result in mass spectra of $\gamma \approx -2$ whereas
those that are mass segregated initially will have slightly steeper
mass spectra $\gamma \approx -2.5$.

It is important to note that the above derivation relies upon all the stars
stopping their accretion at the same time, as the final masses are
a strong function of the time (equation~\ref{maccbhsol}). This should not
be too strong a limitation as in the stellar dominated potential the
accretion timescale must be much longer than the local crossing time
as $\mgas << \mstars$. In a rich young stellar cluster,
gas removal is due to the presence of massive stars. These stars
ionise the gas such that the sound speed ($\approx 10$ km/s)
is comparable to or greater than the velocity dispersion. Thus the
gas should be removed on a timescale shorter than or
comparable to the crossing time, and all the stars in the core
should stop accreting quasi-simultaneously. If this is not the case, then
the resulting mass spectrum will exhibit a local maximum corresponding
to stars that are just outside the cleared region.

In summary, we expect that the combination of a gas dominated and a
stellar dominated regimes and the different accretion physics
operating in each, results in a two power-law IMF. The lower-mass stars
have a shallower ($\gamma \approx -1.5$) slope as their mass accumulation
is dominated by tidal-lobe accretion wheras the higher mass stars have
a steeper ($-2 \ge \gamma\simless -2.5$) slope as their mass accumulation is dominated by
Bondi-Hoyle accretion in the stellar dominated core.

\section{Simulated Cluster IMFs}

In order to explore the relevance of the asymptotic limits for the
mass spectrum, we performed a number of simulations of accretion
onto a cluster of 1000 stars. The stars are initially of equal mass
($0.1 \solm$)
and are embedded in gas which comprises 91 per cent of the total cluster
mass The gas is cold such that it contains 1000 Jeans masses,
\begin{equation}
\label{Jeans_mass}
M_J = \left(\frac{5 R_g T}{2 G \mu}\right)^{3/2} \left(\frac{4}{3} 
\pi \rho \right)^{-1/2},
\end{equation} 
where $\rho$ is the gas density, $T$ is the gas temperature,  $R_g$ is the gas constant, $G$ is the gravitational constant, and
$\mu$ is the mean molecular weight.
The simulations were performed with a hybrid N-body SPH code
(Bate, Bonnell \& Price~1995) which uses standard SPH particles
to model the gas and sink-particles to model the stars (for more
details see also
Bonnell \etal~2000). These 
sink-particles interact only gravitationally with the rest of the
cluster and by accreting gas
particles that come within their sink radius. This sink or accretion
radius was taken to be the smaller of the tidal-lobe radius, the 
Bondi-Hoyle radius or the stellar separation. In addition, gas particles
can only be accreted if they are bound to the star. The simulations were
performed with either 9000 or 90000 SPH particles. The low-resolution
simulations resulted in an excess of stars in the lowest (initial)
mass bins as they were not able to resolve the low accretion rates
involved. The higher-resolution simulations resolved the accretion 
onto all stars and both low and high-resolution simulations resulted
in similar high-mass mass-spectra.

Both the stellar and gas distribution are initially uniform but
collapse down towards a centrally condensed distribution. Gas is
accreted by the stars and removed from the simulation. As the gas is
accreted preferentially near the centre once a power-law density
profile is established ($\rho\propto R^{-2}$), the stars soon dominate
the potential there. The gas must then infall from further out in the
cluster. The simulations result in mass spectra that are broadly
consistent with our analytical expectations. The low-mass stars that
accrete their mass during the gas dominated phase display shallow
($\gamma \approx -3/2$) mass spectra whereas the high-mass stars that
accrete the majority of their mass during the stellar dominated phase
display steeper ($-2.5 \simless\gamma \le -2$) mass spectra. As the
clusters are initially uniform and cold, only the end of the
simulations develop a power-law density profile and that just before
entering the stellar dominated regime. Because of this, the expected
low-mass mass spectrum does not extend over a large mass range.

\begin{figure}
\vspace{-0.35truein}
\psfig{{figure=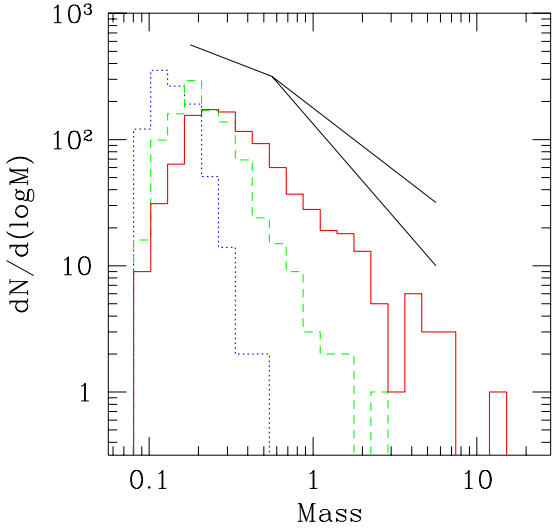,width=3.3truein,height=3.3truein,rwidth=3.0truein,rheight=3.0truein}}
\caption{\label{simimf} Histograms of the resultant mass spectra
are plotted (in log mass bins) for a cluster containing 1000 stars
that initially comprise 10 per cent of the total mass.The mass spectra 
are presented at three different times (dotted line, $t=0.5\tff$; 
dashed line, $t=0.8\tff$; and solid line, $t=0.96 \tff$. The solid 
lines
denote the expected slopes of $\gamma=-3/2$ for low-mass stars
and $\gamma=-2$ and $\gamma=-5/2$ for high-mass stars.}
\end{figure}

In Figure~\ref{simimf}, we plot the resultant mass function for one of
the higher resolution simulations, (performed with 90000 SPH
particles) at three different times during the evolution. The mass
spectrum increases in breadth with time and develops into a power-law
for the high-mass stars. The low-mass mass spectrum is broadly
consistent with tidal-lobe accretion in a mostly uniform cloud,
although only over a small range in mass.  Due to the uniform initial
conditions, only a fraction of the stars experience the power-law
density profile before entering the stellar-dominated phase. This
region would extend further in mass if the cluster was initially
centrally condensed when accretion begins.  At later times, the
high-mass mass spectrum converges to a slope between $\gamma=-2$ and
$\gamma=-2.5$ consistent with the accretion during the stellar
dominated phase, and to the Salpeter slope of $\gamma=-2.35$.

\subsection{Accretion dynamics}

\begin{figure}
\vspace{-0.35truein}
\psfig{{figure=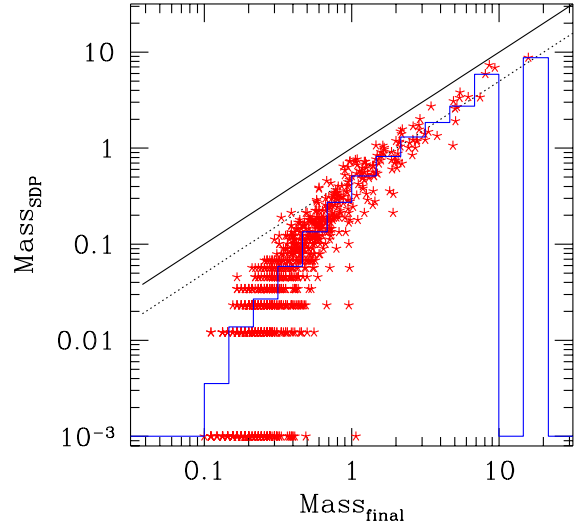,width=3.3truein,height=3.3truein,rwidth=3.0truein,rheight=3.0truein}}
\caption{\label{accdyn} The mass accreted during the stellar-dominated
phase, Mass$_{\rm SDP}$, is plotted against total stellar mass. High
mass stars derive the vast majority of their mass from the accretion in a
stellar dominated potential whereas the majority of low-mass stars
accumulate their mass predominantly in the gas-dominated phase. The
continuous line represents where stars would lie if all their mass
was accreted during the stellar-dominated phase whereas the dotted line
represents where stars would lie if half of their final mass was
accreted during this phase. The solid histogram gives the mean mass
accreted in the stellar dominated potential as a function of final stellar
mass.}
\end{figure}

In order ascertain whether we are correct in our interpretation of a
two-power law IMF resulting from accretion in gas-dominated and
stellar-dominated regimes, we evaluated how much of the eventual
stellar mass was added in each regime. Figure~\ref{accdyn}
plots the amount of mass accumulated by each star during the
stellar-dominated phase against the final stellar mass. High-mass
stars accumulate the majority of their mass during this phase whereas
most of the low-mass stars only accrete a small fraction of their mass
during this phase. The break between the two regimes occurs at
approximately the same final mass ($\approx 1 \solm$) where the mass
function displays a break between the two slopes.  This supports the
assertion that the two different power-laws in the IMF derive from
different physical regimes which affect how the stars accrete. The
low-mass stars derive their mass from tidal-lobe accretion during the
gas-dominated regime whereas the high-mass stars derive the majority
of their mass from a Bondi-Hoyle type accretion that occurs in the
inner parts of the cluster where the potential is dominated by the
stars themselves.

\section{Discussion}

In addition to producing the two power-law mass spectrum, competitive
accretion naturally results in a certain degree of mass segregation.
This arises due to the accretion in the gas-dominated phase where
there is a strong correlation between accretion rate, and thus the
final mass, and position in the cluster (see
equations(\ref{tidmsvsrad}) and (\ref{massvsrad}).  This direct
correlation between the final mass and position in the cluster
neglects variations in the initial masses and the relative movements of
the stars due to their interactions. 
If the cluster is mass segregated entering the stellar dominated phase,
then the $\macc \propto \ms^2$ implies that the mass segregation will
persist. 
Simulations of accretion in clusters show that the mass segregation
does result but that there is not a one to one correlation between
mass and radius (Bonnell \etal~2000).  In fact, low-mass stars are
located throughout the cluster, including in the core, but the high-mass
stars are predominantly located in the central regions as is found in
young stellar clusters such as the ONC (Hillenbrand~1997). 

It is also worth noting that although the models presented here are
meant to consider accretion onto young stars, they are equally
appropriate for the growth of clumps in a molecular cloud. As the
clumps evolve towards gravitational instability, they will accrete
from the surrounding gas and this accretion will be governed by the
physics described here. Thus, for example, the clump mass-function
found by Motte, Andr\'e \& Neri~(1998) for the $\rho$ Oph molecular cloud
could be due to the accretion by the pre-stellar clumps as the whole
system collapses down to form a cluster. The observed $\gamma=-3/2$
slope would imply that the whole system is in a $\rho \propto R^{-2}$
density configuration and is subvirial (dominated by the diffuse  gas
not
in the clumps). The steeper slope found by Motte \etal~(1998) at the
high-mass end of the mass spectrum
can be interpreted as arising from a region of near-uniform
gas density. A test of such a possibility is to
estimate the degree of mass segregation of the clumps in this
pre-stellar cluster system.

Finally, it is possible that the mass spectrum for massive stars, $\ms
\simgreat 10 \solm$, is significantly different than that for lower
mass stars if they do not form in a similar fashion.  If massive stars
cannot accrete above $10 \solm$ due to the effect of radiation
pressure on the infalling dust (Yorke \& Kr\"ugel 1977; Yorke 1993)
but form through a merger process in a dense core (Bonnell \etal~1998)
then the expected mass spectrum could be significantly different from
that presented here.

\section{Conclusions}
Competitive accretion in young stellar clusters results in a two
power-law mass spectrum with a slope of $\gamma\approx - 3/2$ for 
low-mass stars and a steeper slope of $-2 \ge \gamma \simgreat -2.5$ 
for high-mass stars.
The different slopes are due to whether the gas or the stars dominate
the cluster potential. When gas dominates the cluster potential, the
accretion rates are given by a tidal-lobe accretion radius and results
in a $dN \propto \ms^{-{3/2}} d\ms$ mass spectrum in an
isothermal sphere distribution. The shallower density profile in the
central regions implies a steeper mass spectrum, $3/2 \ge \gamma \simgreat -4$.
Once the stars have
accreted enough to dominate the potential, which occurs from the
inside out, the appropriate accretion radius is the Bondi-Hoyle
radius. When this happens the strong dependence of the accretion rate
on stellar mass $\macc \propto \ms^{2}$ results in a 
mass spectrum of $dN \propto \ms^{-{2}} d\ms$ to $dN \propto 
\ms^{-{2.5}} d\ms$.

Simulations of accretion in clusters containing 1000 stars result in
mass spectra broadly consistent with these analytic estimates.  
The high-mass stars are confirmed to accumulate
the majority of their mass in a stellar-dominated potential. 
Lastly, competitive accretion in clusters
naturally results in mass segregation as the accretion rates are
higher near the cluster centre during the gas dominated phase.

\section{Acknowledgements} We thank Hans Zinnecker for useful discussions
and continual enthusiasm. 
IAB acknowledges support from a PPARC advanced fellowship.

\label{lastpage}

\end{document}